\documentclass{article}
\usepackage[utf8]{inputenc}
\usepackage{authblk}
\usepackage{hyperref}
\usepackage{bm}
\usepackage{setspace}
\usepackage{braket}
\usepackage{amsmath}
\usepackage{amssymb}
\usepackage{physics}
\usepackage{mathtools}

\title {Debating the Reliability and Robustness of the Learned Hamiltonian in the Traversable Wormhole Experiment}
\author{Galina Weinstein}
\affil{\normalsize Reichman University, The Efi Arazi School of Computer Science, Herzliya; University of Haifa, The Department of Philosophy, Haifa, Israel.}

\begin{document}

\maketitle

\begin{abstract} 
The paper discusses Daniel Jafferis et al.'s \emph{Nature} publication on "Traversable wormhole dynamics on a quantum processor." The experiment utilized Google's Sycamore quantum processor to simulate a sparse SYK model with a learned Hamiltonian. A debate ensued when Bryce Kobrin, Thomas Schuster, and Norman Yao raised concerns about the learned Hamiltonian's reliability, which Jafferis and the team addressed. Recently, there has been an update in the wormhole experiment saga. In an attempt to rescue the commuting Hamiltonian from its inevitable fate of being invalidated, a recent paper by Ping Gao proposed a creative solution to reinvigorate the concept within the context of teleportation through wormholes. This paper delves into the ongoing debate and the recent endeavor to address the comments made by Kobrin et al. I remain skeptical about the efforts to address Kobrin et al.'s challenges. By its nature, a commuting Hamiltonian does not exhibit chaotic behavior like non-commuting Hamiltonians. Moreover, it's always essential to assess the sensitivity of the Hamiltonian to noise to understand its practical feasibility for the real-world Sycamore processor.
\end{abstract}

\newpage
\section{Introduction}

The experiment of Daniel Jafferis et al. published in \emph{Nature} on "Traversable wormhole dynamics on a quantum processor" \cite{Jafferis} garnered significant attention. The experiment harnessed the computational power of Google's Sycamore quantum processor to simulate a sparse SYK model with the aid of a learned Hamiltonian.\footnote{The Sachdev-Ye-Kitaev (SYK) model serves as a theoretical framework for the onset of quantum chaos and holography. The SYK model was initially proposed in condensed matter physics by Subir Sachdev, Jinwu Ye, and Alexei Kitaev. The model's large $N$ limit exhibits emergent scale invariance, resembling conformal symmetry, due to many $N$ interacting Majorana fermions. As the number of fermions increases, the system displays maximally chaotic behavior, characterized by chaotic dynamics and universal scaling properties of correlation functions. The SYK model demonstrates non-Fermi liquid behavior and possesses a low-temperature entropy that matches the Bekenstein-Hawking entropy of black holes. A notable feature of the SYK model is the all-to-all interactions between Majorana fermions resulting from random and uncorrelated interaction terms. This feature accounts for the model's chaotic behavior and other unconventional phenomena observed in the system.}  This achievement presented exciting prospects for exploring traversable wormholes through quantum means. In Sections \ref{1} and \ref{2}, I introduce the experiment and delve into its intricacies and nuances.

However, as I show in section \ref{4}, the paper's progress was not without its share of controversy.  Bryce Kobrin, Thomas Schuster, and Norman Yao raised valid concerns about the reliability \cite{Galison} of the learned Hamiltonian, prompting a heated debate within the scientific community and popular media \cite{Korbin}. Nevertheless, Jafferis and the team diligently addressed these concerns, attempting to salvage the credibility of their approach \cite{Jafferis1}.
Another paper, authored by Ping Gao, was subsequently published in response to Kobrin et al.'s challenges \cite{Gao2}. Like Jafferis and his research team, Gao attempted to demonstrate that a commuting Hamiltonian is reliable from the standpoint of teleportation through a traversable wormhole. 

In section \ref{4}, I aim to analyze the multifaceted debate surrounding Jafferis et al.'s work and analyze the recent efforts undertaken to confront the critiques put forth by Kobrin et al. I delve into the intricacies of the learned Hamiltonian's reliability and the ingenious proposal that seeks to revitalize the notion of commuting Hamiltonians in the context of holographic wormholes. Through this exploration, I seek to shed light on the current state of affairs in quantum wormhole dynamics and the exciting potential that lies ahead.

In section \ref{5}, I discuss robustness \cite{Galison}. The learned Hamiltonian might lack robustness against noise and errors in the actual conditions in the Sycamore processor, such as fluctuations in gate parameters and environmental noise. As a result, it could produce inaccurate results on the Sycamore processor. I then apply Ian Hacking's notion of material models, which enables scientists to gain insights and improve their understanding of phenomena while suggesting enhancements to experiments, to Jafferis et al.'s experiment. Hacking emphasizes that such models are not exact representations of reality. In light of Hacking's ideas, it is important to question whether the experiment on the Sycamore quantum processor can be regarded as a Planckian wormhole in real physical space. I show that this holds significant implications for the realistic interpretation of the phrase "quantum circuit as wormholes" and calls into question the necessity to simulate the model on a quantum processor.

\section{Teleportation through a traversable wormhole} \label{1}

Adam Brown, Leonard Susskind, and a team of scientists expound in their paper titled "Quantum Gravity in the Lab" that they explore a potential avenue through which quantum gravity can be experimentally examined. Specifically, they investigate a particular entangled state—one that is feasible to create in an atomic physics laboratory — and examine the consequences of introducing a message into the system in a specific manner. One approach to understanding this phenomenon is employing the Schrödinger equation through brute force.
However, the researchers argue that the phenomenon can be elucidated much more straightforwardly within the framework of holographic quantum gravity. They propose an explanation involving a traversable wormhole connecting two black holes, simplifying the understanding of the observed effects \cite{Brown}.

The two black holes or eternal black hole's left and right external bulk regions are linked through a wormhole. They are dual to two identical copies of the original \emph{Conformal Field Theory} (CFT) in the \emph{thermofield double} (TFD) state, as proposed by Juan Maldacena \cite{Maldacena1}. The TFD state represents an entangled pure state between these two copies of the quantum system (CFT):
\vspace{2mm} 

\begin{equation} \label{eq1}
\frac{1}{\sqrt{Z_{\beta}}} \ket{\text{TFD}_{\beta}} = e^{-\beta(H_L+H_R)} \ket{n}_{L} \otimes \ket{n}_{R}, 	  
\end{equation}
\vspace{2mm} 

where $\ket{n}_{L}$ and $\ket{n}_{LR}$ are the energy eigenstates of the right system and the left system, respectively.\footnote{System $R$ is introduced to mimic system $L$ and ensure maximal entanglement between the two systems. We can effectively treat the mixed state as pure by working with the TFD state instead of the thermal density matrix. A thermal state is described by a density matrix: $\rho = e^{-\beta H}$, where $\beta = 1/k_{B T}$ is the inverse temperature, with $k_B$ being the Boltzmann constant and $T$ the temperature. This density matrix represents a mixed state. To purify $\rho$, we introduce an additional Hilbert space that is isomorphic to the original Hilbert space of the system. Instead of working with the thermal density matrix $\rho$, we consider the pure state $\ket \psi$ in the enlarged Hilbert space of the $\ket{\text{TFD}}$. 
In essence, the thermofield double formalism enlarges the Hilbert space and introduces an entangled state that encodes the statistical properties of the thermal state. Let's consider the reduced density matrix of the system L, denoted as $\rho_L$. To obtain $\rho_L$, we trace out (disregard) system R from the TFD state. In other words, we consider subsystem L in isolation and ignore any correlations and entanglement with subsystem R. If we compare $\rho_L$ with the density matrix in $e^{-\beta H_L}$, where $H_L$ is the Hamiltonian of the system L for subsystem L in a thermal state, we can see that they have the same form. When we restrict our attention to subsystem L, the $\ket{\text{TFD}}$ state appears indistinguishable from a thermal state described by $e^{-\beta H_L}$. This similarity implies that the reduced state of subsystem L when the complete system is in $\ket{\text{TFD}}$ resembles a mixed, thermal state described by $e^{-\beta H_L}$.}

The entangled state of the CFT subsystems manifests as a connected geometry in the bulk, represented by the wormhole.\footnote{This connection suggests that entanglement is not merely an abstract concept but has a corresponding geometrical interpretation in the bulk spacetime. A wormhole links the entangled pairs in the CFT. Two copies of the CFT in TFD state, equation (\ref{eq1}), would be dual to the two-sided eternal black hole, and an eternal black hole can be thought of as having two separate black holes connected by a wormhole.}
This dual possibility made Gao, Jafferis, and Aron C. Wall realize they could create a teleportation model through a traversable wormhole \cite{Gao}, \cite{Gao1}. Brown, Susskind, et al. discovered that a process called \emph{unscrambling} comes after scrambling in a wormhole. 

The discovery of the scrambling and unscrambling process has significantly enhanced the possibility of realizing a quantum teleportation protocol in the lab called \emph{teleportation-by-size}: a qubit (message) is scrambled on the left side of the wormhole. Since the two sides, right and left of the wormhole, are coupled, the qubit is unscrambled and pops out on the right side. In the dual gravitational interpretation, teleportation-by-size leads to the interesting conclusion that a particle can pass through the holographic wormhole. 
Two essential things enable traversability: the two sides of the wormhole must be entangled before sending the information, and the two sides must be coupled after sending the message \cite{Brown}. 

Brown, Susskind, and their team introduced an Ansatz known as \emph{size winding} to describe the transmission of a signal through a semi-classical holographic wormhole. Initially, they described size winding from a boundary perspective and then extended its application to the traversable wormholes within the bulk.

Around the \emph{scrambling time} concerning the SYK model, they inserted a thermal operator denoted as $P$ into the left boundary (corresponding to the left side $L$).\footnote{The scrambling time refers to the characteristic timescale within the system during which information becomes highly mixed and spread out throughout the system.} The growth of the operator's size serves as a fundamental manifestation of quantum chaos and the system's complexity. Notably, the distribution of operator sizes exhibits a winding pattern in the clockwise direction. A coupling is applied to examine the influence of coupling between the two subsystems, $L$ and $R$. This coupling leads to the unwinding of the complex winding of the operator size distribution and results in winding the size distribution in the opposite direction, effectively reversing the winding direction. Consequently, the thermal operator $P$ from the left side becomes accurately mapped to its counterpart on the right side. Size-winding behavior in thermal operators near the scrambling time has been demonstrated for the SYK model, and there is a conjecture that this phenomenon could also be present in other holographic systems.

Furthermore, \emph{perfect-size winding} is necessary for traversable wormhole behavior. It specifically occurs in the ground state, where the temperature is at its lowest possible point (essentially zero temperature through the wormhole). In this context, the ground state of a pair of coupled SYK models closely resembles the TFD state. It has been anticipated that systems with a holographic dual would exhibit this perfect size winding behavior \cite{Brown}, \cite{Nezami}.

As a consequence of the \emph{teleportation-by-size} protocol and the \emph{perfect-size winding} Ansatz, a team of researchers led by Jafferis suggested a quantum circuit that draws inspiration from the quantum circuit proposed by Brown, Susskind, and their team for the concept of wormhole teleportation. The researchers aimed to develop a quantum circuit that emulates a holographic wormhole. Specifically, the quantum circuit is designed to be dual to a semi-classical holographic traversable wormhole. The following outlines the primary steps of the circuit\cite{Brown} and \cite{Jafferis2}:

Two identical copies of the quantum system - SYK models - are prepared: a system of $N$ qubits on the left (side $L$) and a system of $N$ qubits on the right (side $R$). The two subsystems are initially entangled in the TFD state.

Register $P$ holds a reference qubit entangled with a qubit $q$ (the message) on register $Q$; both are inserted into the wormhole.\footnote{Several components are involved in the described quantum circuit: registers $P$, $Q$, $L$, $R$, and $T$, SWAP gates, unitary operators, and interaction across the left and right subsystems with coupling operators. Quantum teleportation within a wormhole aims to replicate the teleportation process observed in quantum mechanics. Recall that teleportation involves transferring the state of a qubit from one location to another without physically moving the qubit itself. In teleportation through a traversable wormhole, specifically utilizing the Google Sycamore chip, a reference qubit is needed to establish entanglement with the qubit on register $Q$. The reference qubit allows the transfer of quantum information from the qubit on register $Q$ to the corresponding qubit situated on the opposite side of the wormhole. This process resembles a typical teleportation protocol found in quantum mechanics.}

1) The first step involves evolving all the qubits on side $L$ (register $L$) backward in time by applying the inverse of the time-evolution operator ($\exp^{-iHt}$).\footnote{The notion of "running time backward" does not refer to physically reversing the flow of time. Instead, it is a conceptual approach to describe the sequence of events in the teleportation process through a wormhole. Running time backward in this context means we consider the qubit injection event to occur at a negative time ($-t_0$) relative to the coupling interaction at $t = 0$. This choice of assigning negative values to the time parameter is merely a convention to consistently describe the sequence of events.} A qubit $q$ is injected, namely, swapped into the left side $L$ at the time $t= -t_0$.\footnote{The SWAP gates achieve qubit injection and arrival readout in the teleportation protocol. The SWAP gates enable the qubit on register $Q$ to enter the wormhole and the resulting qubit on register $T$ to be read out at the appropriate times.} 

2) Subsequently, we progress by evolving register $L$ forward in time, utilizing the time-evolution operator ($\exp^{-iHt}$).\footnote{The time evolution operators are implemented using a series of single qubit gates and the controlled $Z$ gate.} Consequently, $q$ becomes rapidly scrambled and entangled with the carrier qubits present in subsystem $L$.

3) Step 3 involves weakly coupling side $L$ to side $R$ at $t=0$, employing the coupling operator $\exp^{i\mu V}$, where the operator $V$ is defined in equation (\ref{Eq9}). 

This coupling occurs suddenly, connecting all the qubits on side $L$ with those on side $R$.\footnote{This interaction, along with the time evolution, is implemented using a Trotter step. The single Trotter step involves breaking down the exponential evolution operator into smaller operations, typically achieved through single-qubit and controlled gates \cite{Jafferis1}, \cite{Jafferis2}, \cite{Zlokapa1}.}
  
4) Finally, in Step 4, we evolve all the qubits on side $R$ forward in time using the time-evolution operator ($\exp^{-iHt}$). At a later time $t=t_1$, we measure side $R$ (register $T$). Remarkably, the qubit $q$ reappears on the side $R$ unscrambled, requiring no decoding. The message is effortlessly refocused on side $R$.

To measure the entanglement of the qubits, one computes the mutual information $I_{PT}(t)$:\footnote{The mutual information $I_{PT}$ measures the correlation shared between the two systems. The $I_{PT}$ tracks the flow of information between the sender and the receiver during this process. It quantifies how much information is gained about the qubit's state as a function of time. A peak in $I_{PT}$ within a certain time window indicates the occurrence of quantum teleportation. This peak represents when the left side obtains the maximum information about the qubit's state, suggesting that the qubit has successfully teleported through the wormhole.}

\begin{equation} \label{Eq8}
I_{PT}(t) = S_{P}(t) + S_{T}(t) - S_{PT}(t),    	  
\end{equation}

\noindent where $S$ is the von Neumann entropy. 

Brown, Susskind et al. and Jafferis et al. identified two distinct mechanisms through which the quantum circuit can accomplish teleportation \cite{Brown},\cite{Jafferis}:

1. \emph{Low-temperature teleportation, near the scrambling time}. 
This regime involves teleportation through the wormhole and is applicable to Hamiltonians with a holographic dual. There are two transmission mechanisms associated with low-temperature teleportation, with a strong dependence on the sign of $\mu$ \cite{Brown}:

a) If $\mu < 0$, the qubit $q$ experiences a time advance and reappears on side $R$, representing traversable wormhole teleportation. In this case, a peak in the signal $I_{PT}(t)$ is reached around $t$ ($t$ being approximately the scrambling time), indicating quantum teleportation within the time window when the wormhole is traversable.
The $LR$ coupling applied between the two sides of the wormhole allows for its traversability. 
If $\mu < 0$, the coupling operator generates a negative energy shockwave in the bulk, modifying the wormhole's geometry and enabling traversability. The traversing qubit experiences a Shapiro time advance upon encountering the pulse of negative energy shockwave, causing it to emerge on the wormhole's other side ($R$) at $t = t_1$.

b) On the contrary, in the case where $\mu > 0$, the qubit becomes entangled with the qubits on side $L$, but it remains scrambled, indicating scrambling teleportation. This scenario involves a reversal of the sign of the coupling interaction, resulting in a positive energy shockwave leading to the generation of a positive energy shockwave that draws the qubit toward the singularity instead of repelling it. Consequently, the qubit is inevitably drawn to the singularity, preventing its emergence on the wormhole's right side ($R$).

2. \emph{High-temperature teleportation, for $t>$ the scrambling time}. 
This mechanism, unexpected from gravity, does not involve signals traversing a geometric wormhole.

Jafferis and his team opted to examine the hypothesis of many-body teleportation by simulating the described quantum circuit on the quantum device called \emph{Sycamore} developed by Google \cite{Jafferis}.

In particular, the SYK model is only dual to teleportation through a semi-classical holographic wormhole in the low-temperature limit, and in this regime, it exhibits perfect-size winding. However, this relationship holds specifically for scenarios involving large $N$ Majorana fermions interacting with a large number ($q$) of other Majorana fermions, effectively allowing for the teleportation of $q$ fermions.
This presented a considerable challenge as simulating the dense SYK model on a noisy quantum device was deemed impractical. In response to the aforementioned findings, the team of scholars led by Jafferis embarked on a project to simplify the SYK model.

\section{Simplification and sparsification} \label{2}

Certain simplifications were made. To study the emergence of gravitational behavior at small $N$ and a few terms of the Hamiltonian, Jafferis and his team considered the Hamiltonian dual to a two-dimensional Anti-de Sitter (AdS2) space. This dual relationship allowed them to relate the properties of the SYK model to gravitational phenomena in AdS2. The Hamiltonian dual to nearly AdS2 space was expressed in terms of the left Hamiltonian ($H_L$) and right Hamiltonian ($H_R$), as well as the Majorana fermions and the coupling constant $J_{ijkl}$ associated with the interaction between the left and right sectors. The left Hamiltonian ($H_L$) and right Hamiltonian ($H_R$) refer to the Hamiltonians of the left and right CFTs, respectively.
In the context of the Hamiltonian dual to nearly AdS2 space:

\begin{equation} \label{Eq1}
H_L = \sum_{1\leq i<j<k<l\leq N} J_{ijkl}\psi_{L}^{i} \psi_{L}^{j} \psi_{L}^{k} \psi_{L}^{l}, 
\quad H_R = \sum_{1\leq i<j<k<l\leq N} J_{ijkl}\psi_{R}^{i} \psi_{R}^{j} \psi_{R}^{k} \psi_{R}^{l}, 
\end{equation}

$\quad J_{ijkl} \sim N(0, \sigma^2)$,
\vspace{2mm} 

\noindent the coupling constant $J_{ijkl}$ is considered a random variable drawn from a Gaussian distribution \cite{Zlokapa}, \cite{Zlokapa1}.\footnote{It is commonly assumed to follow a normal distribution with zero mean and variance of $\sigma^2$:
$J_{ijkl} \sim N(0, \sigma^2)$.
This assumption of randomness in the coupling constant $J_{ijkl}$ arises from the chaotic behavior of the underlying quantum system, which is a key feature of the AdS2/CFT correspondence.}

The large-$N$ SYK model dual to nearly-AdS2 is expected to exhibit perfect-size winding behavior in the low-temperature limit. 

Afterward, Jafferis and his team posed a fundamental question: What is the most straightforward Hamiltonian that preserves the gravitational physics observed in the dense SYK model? Their response to this inquiry is presented in their paper published in \emph{Nature}: "Our numerical simulation shows that $N = 10$ [with $210$ terms] is sufficient to produce such traversable wormhole behavior" \cite{Jafferis}. 
The $N = 10$ SYK model represents a system composed of $10$ interacting Majorana fermions. Jafferis and his team demonstrated that the $N = 10$, $q = 4$ dense SYK model is sufficient to exhibit the size winding behavior and, more broadly, the phenomenon of teleportation through a wormhole. Their findings revealed that the ground state of this model closely resembles a TFD state, highlighting the system's potential for emulating wormhole teleportation. They simulated the $N = 10$ SYK model on a classical computer and collected data that provided insights into perfect-size winding. Once the simulation was complete, the collected data was analyzed to investigate whether the system exhibited the desired teleportation through a wormhole-like behavior in the low-temperature regime.  

But simulating this SYK model on a noisy quantum processor is very challenging due to the complexity of the model and the required number of qubits and gates.\footnote{Simulating the described $N = 10$ SYK model on the Sycamore quantum computer is currently not feasible for the following reasons. The $N = 10$ SYK model requires $10$ Majorana fermions, typically represented using $20$ qubits. Each Majorana fermion corresponds to two qubits. In a quantum processor, qubits are usually represented using two-level quantum systems, such as the energy levels of a superconducting circuit. However, in the case of Majorana fermions, they are represented using two qubits. The two qubits correspond to the two possible states of the Majorana fermion, which are related to each other in a non-local manner. These qubits encode the information about the occupation of the Majorana fermion. The Sycamore processor with $72$ qubits cannot accommodate the required qubits. The SYK model requires a large number of gate operations. However, the Sycamore processor has a limited number of gate operations before the quantum states become too noisy due to errors. Moreover, implementing the complex SYK interactions using the native gate set of the Sycamore processor is not straightforward. As the number of qubits and gates increases, the circuit depth grows, and the susceptibility to noise and errors also increases. The $N = 10$ SYK model would require a substantial circuit depth, which is likely beyond the capabilities of the Sycamore processor. Given these limitations, simulating an $N = 10$ SYK model on the Google Sycamore quantum chip is currently not viable. Hence, for simulating the $N = 10$ SYK model, Jafferis and his research team used classical computational methods.} 

The team subsequently used classical machine learning to reproduce the teleportation behavior of the $N = 10$ SYK model with only a few Hamiltonian terms [see equation (\ref{Eq10}), section \ref{4}].\footnote{A kind of approximation called \emph{Trotterization} was applied to optimize the procedure. Many couplings $J_{j_1 \ldots j_4}$ in equation (\ref{Eq1}) were set to zero to obtain the sparse Hamiltonian. This sparsity was introduced to simplify the model and reduce its computational complexity.}
The sparsification procedure reduced the $N = 10$ SYK Hamiltonian with $210$ terms to a large population of candidate sparse Hamiltonians thought to preserve the gravitational properties of the original model (see equations in section \ref{4}). 

Jafferis and his research team aimed to simplify the $N = 10$, $q = 4$ SYK model by reducing it to a Hamiltonian with the smallest $N$ and the most reduced number of terms while preserving the essential characteristics of the original model, specifically the gravity-like effects. 
They utilized machine learning techniques to acquire a Hamiltonian operator with only a few non-zero terms.\footnote{Using machine learning techniques on a classical computer, the team trained an analog of a neural network to sparsify the SYK model with $N = 10$ and $210$ terms, backpropagated over the Hamiltonian coefficients, and applied regularization, interpreting the Hamiltonian coefficients as neural network weights. Each term in the $N = 10$ SYK Hamiltonian was treated as a feature, and the corresponding coefficient was seen as the weight of that feature. Each data point in the dataset represented a specific configuration of the Hamiltonian terms. A neural network was constructed that mimics the structure of the SYK model. The network had an input layer with neurons corresponding to each term in the original Hamiltonian. These neurons serve as the inputs to the neural network. The training process involved adjusting the neural network's weights (Hamiltonian coefficients) to sparsify the model. This was done using backpropagation, where the errors were propagated from the output layer back to the input layer, and the weights were updated accordingly.
The training objective was to find a set of weights that reduce the $N = 10$ SYK model to an $N = 7$ model with five terms. To encourage sparsity, weight regularization techniques were applied. These techniques introduced a penalty term encouraging the model to have fewer non-zero weights. This helped select the most relevant terms and discard the less significant ones. Terms with low weights were removed. The Hamiltonian coefficients were interpreted as the neural network weights once the training process was complete. The non-zero coefficients indicate the terms selected in the reduced $N = 7$ model.} 

The following $N = 7$ model, referred to as the \emph{learned Hamiltonian}, represents the result consisting of five terms:

\begin{equation} \label{Eq2}
H_{L,R}=-0.36\psi^1\psi^2\psi^4\psi^5+0.19\psi^1\psi^3\psi^4\psi^7 	  
\end{equation}

$-0.71\psi^1\psi^3\psi^5\psi^6+0.22\psi^2\psi^3\psi^4\psi^6+0.49\psi^2\psi^3\psi^5\psi^7.$ 
\vspace{3mm} 

\noindent According to the team's claim, this model successfully captures the fundamental characteristics of the original $N = 10$ SYK model while substantially reducing its overall complexity \cite{Jafferis1}, \cite{Jafferis2}. 

But the sparsification process resulted in an $N = 7$ Hamiltonian with only commuting terms. Since the reduced $N = 7$ learned Hamiltonian consists solely of commuting terms, other scientists suggested that the specific features related to the gravitational behavior were not accurately captured in the simplified model \cite{Korbin}, as discussed in section \ref{4}.

Reproducing the gravitational behavior of a traversable wormhole using the sparse $N = 7$ SYK Hamiltonian is challenging. The success of sparsification depends on various factors, including the details of the machine learning procedure, the quality and quantity of training, and the underlying assumptions and approximations made during the modeling process.
The machine-learning procedure uses a reduced number of Hamiltonian terms to mimic the teleportation behavior through a wormhole of the dense $N = 10$ SYK model. This results in a Hamiltonian, denoted as $H_{L,R}$, involving specific operator combinations. When the teleportation protocol is implemented using this Hamiltonian, which is holographically dual to gravity, the process is described from the perspective of a particle traveling through the wormhole.

As demonstrated by equations (\ref{Eq1}) and (\ref{Eq2}), the Hamiltonian is doubled to yield left $H_L$ and right $H_R$ Hamiltonians, each simulating the SYK model and each containing seven Majorana fermions on their respective sides. In addition, the wormhole teleportation protocol incorporates a pair of entangled qubits, namely, a reference qubit that is entangled with the injected qubit. Consequently, the entire circuit involves a total of nine qubits. The five terms in the Hamiltonian describe the interactions between these qubits, forming the $9$-qubit circuit for the protocol \cite{Jafferis}.\footnote{Considering two SYK models, each with $N = 7$, and the need for two additional qubits for teleportation, the total number of qubits used on the Google Sycamore chip would be seven qubits for the left SYK model, seven qubits for the right SYK model, and two qubits for the teleportation process. Therefore, in this scenario, the total number of qubits used on the Sycamore chip would be $7 + 7 + 2 = 16$ qubits.}  

To determine the ground state of the learned Hamiltonian (equation \ref{Eq2}), the researchers employed a hybrid classical-quantum algorithm known as the \emph{Variational Quantum Eigensolver} (VQE). They utilized the VQE algorithm to the following Hamiltonian:\footnote{Initializing the wormhole at $t = 0$ with $\ket{\text{TFD}}$ using a VQE algorithm is done on the Google Sycamore chip by preparing the individual qubits in the state $\ket{0}$, and then entangling them appropriately using a sequence of controlled gates and single-qubit rotations.} 

\begin{equation} \label{Eq6}
H_{TFD}= H_L + H_R + i\mu V, 	  
\end{equation}

\noindent where $V$ is the interaction term:

\begin{equation} \label{Eq9}
V=\sum_j \psi{^j_L}\psi{^j_R}, 	  
\end{equation}

\noindent and $\mu$ represents the coupling interaction. 

The researchers' findings reveal that the ground state of $H_{TFD}$ closely approximates the TFD state.

Equation (\ref{Eq6}) bears a resemblance to the Hamiltonian corresponding to the traversable wormhole equivalent, known as the \emph{eternal traversable wormhole Hamiltonian}, denoted as $H_{tot}$:

\begin{equation} \label{Eq5}
H_{tot}= H_L + H_R + i\mu' V. 	  
\end{equation}

This similarity is a consequence of the ground-state behavior of the coupled Hamiltonian.

The team of scholars demonstrated that the learned Hamiltonian, represented by equation \ref{Eq2}, displays similar scrambling and thermalization behavior to the original $N = 10$ SYK model. This behavior is characterized by the two-point and four-point functions, where the two-point function indicates thermalization time, and the four-point function indicates scrambling time.\footnote{The two-point function is a correlation function that tracks the relationship between two observables as time progresses in a quantum system. In thermalizing quantum systems, correlations between observables diminish over time until thermal equilibrium is reached, and this duration is the thermalization time. The four-point correlation function probes scrambling behavior in chaotic quantum systems. In such systems, the four-point function deviates significantly from its initial value and rapidly increases, indicating the spread of entanglement, and the time at which it reaches a characteristic value is referred to as the scrambling time.}

In the left side $H_L$, the transmitted fermions undergo thermalization and scrambling, as evidenced by the decay of two-point correlators and out-of-time ordered correlators (OTOC). The averages of these correlators over Majorana operators were plotted, demonstrating a correspondence between the curves of the $N = 10$ SYK model and the learned Hamiltonian \cite{Jafferis}, \cite{Jafferis2}.

It is worth noting that the large-$N$ SYK model exhibits a significant difference between the thermalization time and the scrambling time, with the scrambling time occurring just before the onset of chaotic behavior. However, the scrambling time is approximately equal to the thermalization time for the $N = 10$ SYK model and the learned Hamiltonian represented by equation (\ref{Eq2}). This observation is consistent with wormhole-like teleportation properties.

Jafferis et al. first investigated the low-temperature limit and verified that the behavior of the learned Hamiltonian aligns with the original $N = 10$ SYK model for both $\mu = -12$ and $\mu = +12$. This correspondence was illustrated through the similarity in the curves of the two models. Notably, the similarity was evident in the peak positions of the $N = 10$ SYK model and the learned Hamiltonian [see equation \ref{Eq2}].
For a fixed injection time of $-t_0 = -2.8$, it was anticipated that $I_{PT}(t)$ would reach its peak around the scrambling time $t_*$ ($t_0 \approx t_1 \approx t_*$) for $\mu = -12$ (the interaction) due to the negative energy shockwave making the wormhole traversable. Through a simulation carried out on the Sycamore quantum processor with gate errors, a peak in $I_{PT}(t)$ was observed solely for $\mu = -12$, suggesting the occurrence of teleportation.

The outcomes also revealed an asymmetry in $I_{PT}(t)$ when considering opposite signs of $\mu$, specifically, a peak for $\mu = -12$ and a trough for $\mu = +12$ (corresponding to teleportation occurring by means of scrambling instead of through the wormhole, consistent with theoretical expectations). This observation demonstrated qualitative consistency with the numerical simulation performed on classical computers using the $N = 10$ SYK model.

Furthermore, the researchers demonstrated that in the high-temperature regime, non-gravitational teleportation occurred, and there was no size-winding effect \cite{Jafferis}.

Multiple noisy simulations were conducted, and a noteworthy trend emerged regarding the behavior of traversable wormhole teleportation. It was consistently observed that $I_{PT}(t)$ exhibited a peak around the scrambling time across various noisy runs. On the other hand, when coherent noise was present, it was more likely to lead to random fluctuations in $I_{PT}(t)$. The above signatures were verified on classical computers, confirming that the quantum system dynamics were consistent with a quantum gravity interpretation and the holographic principle \cite{Zlokapa}.
Based on these findings, Jafferis and his team concluded:  Teleportation was more pronounced when the interaction introduced a negative energy shockwave than a positive one. The distinct asymmetric signature observed in the results aligns with the physical interpretation that the qubit underwent teleportation through the wormhole \cite{Jafferis}. 

Graphs were presented that show curves with their coarse-grained SYK model preserving key properties of the traversable wormhole physics: perfect size winding, coupling interaction on either side of the wormhole that is consistent with a negative energy shock wave, a Shapiro time delay, causal time-order of signals emerging from the wormhole, and scrambling and unscrambling \cite{Jafferis}.\footnote{In addition to teleporting a single qubit from the left side to the right side, the researchers conducted another experiment where a qubit was sent from the right side to the left side. This resulted in what they termed "time-ordered teleportation," which they interpreted as an indication of gravitational teleportation. The process involved swapping a qubit $Q$ into the left side ($L$) at time $-t_0$, and simultaneously swapping a qubit $R$ into the right side ($R$). At the time $t_1$, the team performed measurements and compared the two processes.
In this experiment, the causal time ordering played a crucial role. Qubits inserted earlier emerged while qubits inserted later would pop out later, providing evidence supporting the gravitational interpretation.
However, sending one qubit into $L$ and another into $R$ with the expectation of them meeting in the middle resulted in a slight delay. This phenomenon was identified as a "Shapiro time delay," wherein a qubit injected into $R$ would interact with a qubit injected into $L$ \cite{Jafferis}, \cite{Zlokapa1}.} 

\newpage
\section{Examining the controversy: the reliability of the learned Hamiltonian} \label{4}

A few months after the publication of Jafferis et al.'s \emph{Nature} paper, a comment was published by three researchers led by Norman Yao. Their comment raised concerns about the validity of Jafferis et al.'s $5$-term commuting learned Hamiltonian, as described in equation (\ref{Eq2}). They identified multiple flaws and challenged its reliability \cite{Korbin}. 
Kobrin et al. focus on evaluating the reliability of the sparsification method and the learned Hamiltonian. 
The central aim of Yao and his team's inquiry was to establish whether this equation aligns with the gravitational dynamics (specifically, a qubit emerging from a traversable wormhole) of the original SYK model. 
In other words, they were interested in determining whether the Hamiltonian could be considered a dependable model for simulating the teleportation of a qubit through a semi-classical wormhole \cite{Weinstein}. 

Recall from section \ref{2} that certain simplifications were made to the dense $N = 10$ SYK model during the sparsification process to reduce its complexity. These idealizations were instrumental in obtaining a sparser representation of the model while preserving its essential gravitational properties. However, sparsification comes with a trade-off between model complexity and accuracy. While a simpler model may be suitable for implementation on the Google Sycamore quantum processor, it may not fully capture all the intricate details of the original dense SYK model. Some physical effects might be overlooked or underestimated in the process.
Additionally, the sparsified Hamiltonian has limitations in its range of validity. Its predictions and behavior may only be reliable within certain conditions. To address these potential issues, Jafferis et al. carefully analyzed the reliability and applicability of the sparsified learned Hamiltonian in their experiment. They demonstrated that the behavior of the mutual information $I_{PT}(t)$ between the $N=10$ model and the learned Hamiltonian remains almost the same, indicating that the sparsified model captures relevant information despite its simplifications.
By thoroughly examining the agreement between the sparsified learned Hamiltonian and the $N = 10$ SYK model, they sought to validate the reliability of their simplified representation within the experimental setup. 

But some critics argued that the measures taken were insufficient to tackle the issues adequately. Kobrin et al. discovered a gap in equation (\ref{Eq2}), asserting that it fails to encompass both the global physics seen in typical SYK models and the complete characteristics of gravitational physics. In other words, the equation does not fully represent the broader aspects of ordinary SYK models nor adequately capture the intricacies associated with gravitational physics.

According to Kobrin et al., the learned Hamiltonian falls short in representing the dense SYK model in the following aspects \cite{Korbin}:

1. The teleportation signal only resembles the $N = 10$ SYK model used in the machine-learning training process for certain Majorana operators. The complexity of the teleportation through the wormhole and the associated dynamics in the $N = 10$ SYK model is influenced by the interactions among different operators and the specific Hamiltonian terms. While effective for the trained operators, the machine-learning procedure may not fully capture the intricacies of all operators in the system, leading to limited resemblance to the behavior of the $N = 10$ SYK model. 
During the machine-learning procedure, the training is conducted on teleportation involving specific fermions, namely $\psi^1$ and $\psi^2$. Jafferis and his team evaluate the interaction on these two qubits numerically, involving the time evolution operator $\exp^{iHt}$, which yields results demonstrating asymmetry for opposite signs of $\mu$ (where $\mu=12$ represents scrambling teleportation and $\mu=-12$ represents wormhole teleportation). This finding indicates a qualitative agreement of equation (\ref{Eq2}) with the numerical simulation of the $N = 10$ SYK model.

However, Kobrin et al. argue that the observed teleportation signal, characterized by a negative energy shockwave, causal time-ordering of teleported signals, and a Shapiro time delay, as well as perfect size winding, is specific to the fermions $\psi^1$ and $\psi^2$, which were involved in the training process. The agreement between equation (\ref{Eq2}) and the $N = 10$ SYK model is observed for this trained pair of fermions. Yet, these distinctive characteristics are not observed when the teleportation protocol is attempted with other fermions not included in the training procedure.

Kobrin et al. specifically identified an issue with the untrained fermions $\psi^4$ and $\psi^7$. They observed that these fermions display poor size winding at $t_0\approx 2.8$, which is the time of the wormhole teleportation signal or injection time. In simpler terms, the size winding and teleportation behavior, involving $\mu = -12$ and at $t_0\approx 2.8$, are exclusively observed for the trained Majorana fermions $\psi^1$ and $\psi^2$, and not for any other fermions that were not part of the training process.

To generate the Hamiltonian, the machine-learning procedure is specifically trained on the behavior of two operators, $\psi^1$ and $\psi^2$. Consequently, the Hamiltonian is optimized to capture the teleportation dynamics involving these specific operators. However, its performance might not generalize well to other operators, such as $\psi^3$ and $\psi^4$, as the Hamiltonian is tailored to the training data of $\psi^1$ and $\psi^2$. Thus, the behavior of untrained operators might not be accurately represented in the generated Hamiltonian.

2. Equation (\ref{Eq2}) comprises solely commuting terms, but a crucial distinction exists in the structure of time-evolved operators between fully-commuting models, such as equation (\ref{Eq2}), and non-commuting models, like the original SYK model.
The original SYK model, with non-commuting Hamiltonian terms, exhibits distinctive properties such as information scrambling, strong entanglement, rapid thermalization, and chaotic behavior. The non-commutative nature of the Hamiltonian terms plays a crucial role in generating these characteristics. In this model, interactions between Majorana fermions occur all-to-all, allowing any pair of fermions to interact.
On the other hand, if we consider a fully commuting Hamiltonian, where each term commutes with every other term, the system demonstrates more regular and integrable behavior. As acknowledged in their paper, Jafferis and his research team state that the Hamiltonian in question with five terms consists of all commuting terms \cite{Jafferis1}.

Kobrin et al. examine two alternatives to equation (\ref{Eq2}): 

1) In Jafferis et al.'s \emph{Nature} paper, they demonstrate that the machine-learning procedure used to derive equation (\ref{Eq2}) actually generates a $6$-term Hamiltonian. Unlike equation (\ref{Eq2}), this $6$-term Hamiltonian does not consist entirely of commuting terms \cite{Jafferis}:

\vspace{1mm} 
\begin{equation} \label{Eq3}
H_{L,R}=-0.35\psi^1\psi^2\psi^3\psi^6+0.11\psi^1\psi^2\psi^3\psi^8 	  
\end{equation}

$-0.17\psi^1\psi^2\psi^4\psi^7-0.67\psi^1\psi^3\psi^5\psi^7+0.38\psi^2\psi^3\psi^6\psi^7-0.05\psi^2\psi^5\psi^6\psi^7.$ 
\vspace{5mm} 

\noindent Equation (\ref{Eq3}) represents a weakly perturbed version of a fully commuting Hamiltonian. 

However, Kobrin et al. identified a flaw in this Hamiltonian. They demonstrate that the key observations made about equation (\ref{Eq2}) are also applicable to equation (\ref{Eq3}). In other words, the issues and concerns raised in relation to equation (\ref{Eq2}) carry over to the perturbed equation (\ref{Eq3}).
Specifically, the teleportation signal does not resemble the $N = 10$ SYK model for untrained operators. However, the size winding behavior is similar to that observed in equation (\ref{Eq2}). Notably, there is a distinct difference between equation (\ref{Eq3}) and the $N = 10$ SYK model regarding the timescale during which $\psi^1$ and $\psi^2$ were trained. For equation (\ref{Eq3}), the teleportation signal displays an initial peak, followed by significant revivals as time progresses, unlike the $N = 10$ SYK model.

Jafferis and his team assert that "the learned Hamiltonian scrambles and thermalizes similarly to the original SYK model as characterized by the four- and two-point correlators" \cite{Jafferis}. However, Kobrin et al. present a different perspective. Both the SYK model and equation (\ref{Eq2}) display decay in their two-point functions. In the case of the SYK model, this decay aligns with the expected quantum thermalization. However, for equation (\ref{Eq2}), the individual two-point correlators exhibit strong revivals as time progresses. According to Kobrin et al., this behavior suggests that the training procedure was unsuccessful and unreliable. They also emphasize that such discrepancies were not demonstrated or discussed in the \emph{Nature} paper. Consequently, they concluded that the apparent agreement between equation (\ref{Eq2}) and the SYK model is, in reality, an artifact resulting from averaging over the two-point and four-point correlation functions \cite{Korbin}.

2) In their \emph{Nature} paper, Jafferis et al. employ an alternative machine-learning approach to create another Hamiltonian featuring $N=10$ and $8$ terms. The specific objective of this Hamiltonian is to maximize the distinction in the teleportation signal between $-\mu$ (associated with wormhole teleportation) and $+\mu$ (related to scrambling teleportation) \cite{Jafferis}.

\begin{equation} \label{Eq10}
H_{L,R}=0.60\psi^1\psi^3\psi^4\psi^5+0.72\psi^1\psi^3\psi^5\psi^6 	  
\end{equation}

$+0.49\psi^1\psi^5\psi^6\psi^9+0.49\psi^1\psi^5\psi^7\psi^8$
\vspace{1mm} 

$+0.64\psi^2\psi^4\psi^8\psi^{10}-0.75\psi^2\psi^5\psi^7\psi^8+0.58\psi^2\psi^5\psi^7\psi^{10}-0.53\psi^2\psi^7\psi^8\psi^{10}.$ 
\vspace{3mm} 

Jafferis and his team explain that as the number of terms increases in equation (\ref{Eq10}), the number of gates needed for Trotterization also rises linearly. This implies that the gate count required to implement the Hamiltonian must at least double. Unfortunately, the fidelity of circuits exponentially decreases with the number of gates and qubits, and the experimentally measured fidelity was already below half of the noiseless fidelity. Consequently, Jafferis and his team conclude that equation (\ref{Eq10}) "cannot provide a stronger teleportation signal when experimentally measured" and does not exhibit perfect size winding but rather a slightly damped one \cite{Jafferis}.

However, despite the technical explanation being essentially correct, there are dissenting opinions, as Kobrin et al. provide an alternative interpretation, leading to different perspectives on the matter. Kobrin et al. identify a concern with small-size fully-commuting Hamiltonians consisting of only a few terms, including equation (\ref{Eq10}). Despite being non-commuting, they comment that this particular Hamiltonian shows clearer indications of thermalization over a longer time scale, approximately $t\approx 30$. Unlike the expected perfect-size winding, the teleportation signal for this Hamiltonian displays a single peak structure for nearly all operators.
The phenomenon of perfect-size winding is generally observed in fully commuting Hamiltonians with only a few terms. However, it does not persist in larger fully-commuting or non-commuting systems. Kobrin et al. demonstrate that introducing random numerical coefficients in front of the terms in equation (\ref{Eq2}), or considering random commuting terms, also yields a perfect-size winding. Thus, Kobrin et al. conclude that perfect-size winding in small-size fully-commuting Hamiltonians with only a few terms is likely a side effect of the sparsification method and does not imply that equation (\ref{Eq2}) is holographically dual to gravity \cite{Korbin}.

Kobrin et al. conclude that the reported perfect-size winding in the \emph{Nature} paper is contingent on the small size of the system and equation (\ref{Eq2}). 
Therefore, unlike the $N = 10$ SYK model, equation (\ref{Eq2}) does not undergo thermalization, and any agreement in thermalization behavior between the two is merely an artifact \cite{Korbin}. In other words, the observed similarity in thermalization is not a genuine characteristic of equation (\ref{Eq2}), but rather a consequence of the specific conditions in which it was examined.

Kobrin et al. are raising concerns that the machine-learning procedure used by the authors may not have been entirely reliable and unbiased. In essence, they question the validity and robustness of the machine-learning procedure, implying that the obtained results might not be entirely trustworthy.
In other words, the assumption of perfect-size winding in the learned Hamiltonian (for $N=7$) may not be accurate due to the influence of bias introduced during the machine-learning process. They argue that introducing bias among the trained operators could have influenced the results, potentially leading to inaccuracies or misleading conclusions \cite{Korbin}. The presence of this bias may have affected the results, casting uncertainty on the validity of the observed perfect-size winding in relation to teleportation through the wormhole.
Their main point is that the observed perfect-size winding in \cite{Jafferis} is contingent on the small size of the system, which is the fundamental aspect they emphasize. Moreover, these fully-commuting models contradict other essential features of holography, thermalization, complexity, and chaos. This raises doubts about whether the observed perfect-size winding is genuinely connected to the gravitational picture in a significant way.

The criticism focuses on the reliability and appropriateness of the fully-commuting Hamiltonians in reproducing a dual gravitational behavior. The gravitational picture itself is not being questioned or challenged; rather, the scrutiny is directed toward whether these Hamiltonians can effectively model the well-established features of holography that are assumed to be true. In essence, the criticism is centered on the capability of the learned Hamiltonian equation (\ref{Eq2}) to accurately capture the expected holographic properties, given the specific nature of the fully-commuting Hamiltonians under consideration.

One month later, Jafferis and his team published a comment in response to the concerns raised by Kobrin et al. In their response, they presented a compelling solution to defend their learned Hamiltonian, equation (\ref{Eq2}), as expected. Jafferis et al. found an elegant way to counter the main arguments put forth by Kobrin et al.
To properly examine Jafferis et al.'s response, we must analyze its specific points and arguments. It may shed light on how they addressed the criticisms raised by Kobrin et al. and how they defended their learned Hamiltonian, equation (\ref{Eq2}). A thorough examination of their response could provide a clearer understanding of their perspective and whether they could effectively address the concerns raised by Kobrin et al.

The comment argued that the single-sided Hamiltonian $H_L$ in equation (\ref{Eq2}) consists of commuting terms and does not thermalize later after teleportation due to the recurrence of these commuting terms. Jafferis et al. found a way, in the context of the eternal traversable wormhole Hamiltonian $H_{tot}$ to rebut this claim: evolving under $H_{tot}$ [equation (\ref{Eq5})] from the \emph{Nature} paper, shows that equation (\ref{Eq5}) exhibits operator growth and thermalizes at high temperatures after teleportation. A single Trotter step of time evolution under the Hamiltonian $H_{tot}$ can be understood as being equivalent to the teleportation quantum circuit described in the \emph{Nature} paper \cite{Jafferis1}.

Jafferis et al. further found another counterclaim stating the following: In the context of the gravitational interpretation of thermalization rates, each fermion corresponds to a distinct mass. When a fermion is inserted onto the TFD and then time-evolved under a single-sided Hamiltonian $H_L$, it leads to operator growth, where lighter fermions experience slower growth. This phenomenon is attributed to the fact that the wave packet of a lighter fermion is more spread out, causing its two-point function to decay at a slower rate.
Jafferis et al., in their work \cite{Jafferis1}, reach the conclusion that the two-point function described by equation (\ref{Eq5}) decays for all fermions, signifying thermalization. As a result, they note that the fermions exhibiting the slowest operator growth in $H_L$ are precisely the same fermions that thermalize the slowest in $H_{tot}$, as indicated by equation (\ref{Eq5}). In equation (\ref{Eq5}), Jafferis et al. observed that all fermions exhibit significant size winding within the time range $2\gtrsim t \lesssim 5$. On the other hand, Kobrin et al.'s comment only focused on analyzing size winding at a specific time point, approximately $t_0 \approx 2.8$, and claimed it to be an artifact.
Jafferis et al. examined a counterfactual scenario to gain deeper insights, revealing meaningful gravitational behavior. They demonstrated that even at later times, specifically in the range $5\gtrsim t \lesssim 10$, the wormhole teleportation phenomena persist despite the introduction of a strong non-commuting perturbation represented by equation (\ref{Eq4}). Although the late-time dynamics are now governed by a noncommuting Hamiltonian, the system's behavior during teleportation at $t_0 \approx 2.8$ remains unchanged and consistent with the expected gravitational signature \cite{Jafferis1}.

Kobrin et al. assert that equation (\ref{Eq2}) displays a bias, resulting in favorable size winding behavior only for the two fermions implemented in their experiment \cite{Korbin}.
Furthermore, they stress that particular emphasis is placed on a specific pair of input operators, namely, $\psi^1$ and $\psi^{2}$. In other words, Jafferis et. all discovered a representation within the sparse $N = 7$ Hamiltonian that replicated the gravitational behavior demonstrated by the $N = 10$ model when these specific operators were considered. The $N = 7$ Hamiltonian mimics the dynamics and properties observed in the $N = 10$ model, specifically concerning the behavior of these selected operators, $\psi^1$ and $\psi^{2}$.
However, Jafferis and his team challenge this claim and present a compelling physical justification for their findings. They observe that fermions exhibiting slower thermalization in the eternal traversable wormhole Hamiltonian [equation (\ref{Eq5})] demonstrate significant size winding at later times. This behavior is attributed to the varying masses of the fermions: lighter fermions undergo slower thermalization and consequently take longer to traverse the wormhole. Specifically, fermions $\psi^4$ and $\psi^7$ are the slowest to thermalize, leading to their achievement of size winding at slightly later times.

The argument put forth by Jafferis and his team may initially seem peculiar, as it relies on physical reasoning. In contrast, the criticism raised by Kobrin et al. concerning the two operators, specifically fermions $\psi^3$ and $\psi^4$, is grounded in considerations derived from machine learning and sparsification techniques.
Jafferis and his team's explanation is rooted in the physical behavior of the system, where they observe that fermions with slower thermalization achieve size winding at later times, corresponding to different masses across fermions. This reasoning delves into the fundamental physics underlying the phenomena.
On the other hand, Kobrin et al.'s criticism revolves around machine learning and sparsification aspects. They point out that the machine-learning procedure, focused on specific operators such as $\psi^1$ and $\psi^2$, might not effectively generalize to untrained operators like $\psi^3$ and $\psi^4$. This limitation could affect the accuracy of the generated Hamiltonian in capturing the behavior of these untrained fermions during teleportation.
Therefore, these two arguments stem from different perspectives, with Jafferis and his team's approach drawing on physics-based reasoning. In contrast, Kobrin et al.'s critique centers around the machine learning and sparsification aspects of the problem.

Jafferis and his team further point out that the fermions identified in Kobrin et al.'s comment as having poor size winding at $t_0 \approx 2.8$ also experience slower thermalization and operator growth. However, their size winding occurs later at around $t_0 \approx 4$ instead of the previously assumed time of $t_0 \approx 2.8$ as mentioned in the comment. Jafferis and his team interpret this delay as being consistent with the notion that these fermions take longer to traverse the wormhole.
In conclusion, Jafferis and his team establish that, eventually, all fermions achieve size winding, regardless of initial differences, indicating a consistent pattern of behavior in the system \cite{Jafferis1}.

As a reminder, Kobrin et al. discovered an issue with small-size fully-commuting Hamiltonians that consist of only a few terms. They argued that the perfect-size winding observed in those systems is likely a side effect and does not indicate a holographic duality to gravity \cite{Korbin}. Jafferis et al.'s next rebuttal against Kobrin et al.'s arguments concerns the assertion that the commuting structure of equation (\ref{Eq2}) is not directly linked to the size winding properties and dynamics.
Jafferis and his team introduced a significant non-commuting term, denoted as $H_1$, to the original equation (\ref{Eq2}), which they refer to as $H_0$. Therefore, the perturbed Hamiltonian can be expressed as follows:

\begin{equation} \label{Eq4}
H_0 + H_1 = H_0+0.3\psi^1\psi^2\psi^3\psi^5. 	  
\end{equation}

Jafferis and his team straightforwardly addressed Kobrin et al.'s issue by proposing that equation (\ref{Eq4}) exhibits sufficient size winding, leading to an asymmetric teleportation signal. According to Jafferis et al., introducing the non-commuting term, $H_1$, generates a behavior that resembles that of the SYK model. Despite thermalizing later, the perturbed Hamiltonian exhibits comparable size winding and teleportation behavior with a parameter value of $\mu = -12$ and occurs approximately at $t_0 \approx 2.8$.
Jafferis et al. found that the non-commuting perturbation $H_1$ has little impact on the physics during teleportation, indicating that the commuting structure is not essential for the presence of gravitational physics.
Furthermore, similar to the $N = 10$ SYK model, the two fermions, $\psi^1$ and $\psi^2$, traversing the wormhole do not display significant revivals over time after the initial peak. This observation supports their argument and suggests that the behavior exhibited by equation (\ref{Eq4}) aligns with the expected gravitational physics \cite{Jafferis1}.

Jafferis et al. believed that they had effectively addressed the concern raised by Kobrin et al. They thought they had successfully resolved the issue.

As previously mentioned, Kobrin et al. suggested that size winding might result from the sparsification process rather than an inherent characteristic of the Hamiltonians. Moreover, the perfect-size winding is commonly observed in fully commuting Hamiltonians at smaller system sizes \cite{Korbin}. 

In response to this criticism, a recent paper by Gao proposed that a large-$N$ commuting SYK model, with $q = 4$, demonstrates "pseudo-holographic" features, including size winding at high temperatures and a teleportation mechanism that exhibits several similarities to the semiclassical traversable wormhole teleportation but operates in distinct parameter regimes. 
The paper also argued that a commuting SYK model with $N = 8$, $q = 4$, and six terms also displays size winding, namely, that the small $N$ Hamiltonian is nearly identical to the large $N$ case. Gao writes that upon closer examination of his model and the learned Hamiltonian, it becomes apparent that the underlying mechanism is not solely attributed to size winding. Remember that Kobrin et al. proposed a Hamiltonian specifically optimized for achieving the best teleportation fidelity for a particular operator $\psi_1$. However, for other operators $\psi_i$, the quality of size-winding diminishes. In contrast, Gao comments that his $N = 8$ model performs equally well for all operators $\psi_i$, exhibiting a consistent and average level of teleportation efficiency. Furthermore, he claims that the thermalization of his Hamiltonian is nearly complete \cite{Gao2}.

As mentioned in Sections \ref{2} and \ref{4}, high-temperature teleportation occurs for values of $t$ greater than the scrambling time, this mechanism, unexpected from gravity, does not involve signals traversing a geometric wormhole. Finally, Jafferis et al. demonstrated that in the high-temperature regime, non-gravitational teleportation occurred, but there was no size-winding effect \cite{Jafferis}. In fact, Gao points out that upon closer examination of his model, the size winding in the large $N$ limit differs significantly from the size winding in an ordinary SYK model \cite{Gao2}. As such, it remains unclear how Kobrin et al.'s criticism is refuted in light of this current state of affairs. 

There appears to be a complication. Gao notes that the size winding observed in the large $N$ model differs significantly from that in a typical SYK model. The operator size distribution in his model is narrowly peaked, indicating that it does not undergo rapid scrambling like a holographic model would \cite{Gao2}. Rapid scrambling is associated with chaotic behavior. Now, when Gao mentions that the operator size distribution in his model is "narrowly peaked" and "does not scramble as fast as a holographic model," he is suggesting that the information in his model does not spread and get entangled as quickly as it would in a holographic model, such as one with properties analogous to a semi-classical holographic wormhole. Instead, the information remains more localized.
In other words, Gao's model possesses "pseudo-holographic" (or non-holographic) features and has different properties than the holographic models. 

I, therefore, remain skeptical about the large-$N$ and the $N = 8$ commuting Hamiltonians. Given that the large $N$ model does not display significant scrambling and chaotic behavior, an $N = 8$ model derived from it would likely also share similar characteristics. Reducing the number of degrees of freedom from a large $N$ model to $N = 8$ simplifies the system and makes it more tractable. But it seems to me that this reduction also leads to a loss of some complex behaviors observed in the original large $N$ model, including rapid scrambling and chaotic dynamics. In this case, the $N = 8$ model is not expected to show rapid scrambling and chaotic dynamics.

Kobrin et al.'s criticism is based on the notion that it is unlikely for a commuting Hamiltonian at larger system sizes to exhibit perfect-size winding, similar to what is expected from a semi-classical traversable wormhole arising from a non-commuting Hamiltonian at low temperatures. The key point here is that the semi-classical traversable wormhole, associated with large-$N$ non-commuting Hamiltonians in the context of holography, involves fast scrambling and non-trivial quantum dynamics, which are typically absent in commuting systems known for their lack of complexity and chaotic fast scrambling.
In general, a commuting Hamiltonian cannot exhibit chaotic behavior, at least not in the same way as non-commuting Hamiltonians do, because the system's time evolution is integrable. 

\section{Assessing reliability in noisy quantum systems} \label{5}

Jafferis and his team's quantum circuit was run on the Google Sycamore processor, which is susceptible to noise. Indeed, Jafferis et al. conducted a simulation involving inherent noise on the Sycamore quantum processor.

As said in section \ref{2}, Jafferis et al. simplified the $N = 10$ SYK model using machine learning techniques while retaining its crucial gravitational properties. They implemented the sparse model through a quantum circuit that employed $164$ controlled $Z$ gates \footnote{A two-qubit gate introduces entanglement between the qubits, i.e., operates on two qubits simultaneously.} and $295$ single-qubit gates.\footnote{A gate that allows for individual qubit operations.} 
While a more complex model than the learned Hamiltonian could have offered enhanced accuracy, it would also entail a higher number of gates and, consequently, a higher error rate.
To enhance the accuracy of the simulation, Jafferis et al. conducted XEB calibration (Xmon's Error Model and Benchmarking). They addressed readout errors\footnote{Inaccuracies in measuring the quantum state of qubits.}, specifically focusing on the $9$ qubits region of the Sycamore quantum processor.
The Sycamore chip comprises $72$ qubits; however, not all qubits exhibit the same levels of noise and errors. Hence, Jafferis et al. carefully selected the $9$ least noisy qubits on the chip for their experimentation. They then applied the calibration tools to these $9$ qubits.

Despite the inherent errors and noise present in the Sycamore chip, Jafferis et al. successfully achieved a teleportation signal on the 9 noisy qubits, as reported in their \emph{Nature} paper. In their paper, Jafferis et al. reported that their approach, which involved using machine learning to simplify the quantum model while preserving crucial gravitational properties and combining it with calibration techniques, enabled the successful execution of quantum teleportation. According to Jafferis et al., their method proved effective even on qubits characterized by significant noise levels.

However, skeptics could challenge the experiment by proposing that coherent errors cause the observed teleportation signal. As said above, the Sycamore processor is prone to errors and can introduce noise and uncertainty into the experimental results, making it difficult to discern the true quantum teleportation signal.
In their \emph{Nature} paper, Jafferis and his team rebut this claim. They have reported their findings, stating that "device noise is primarily influenced by an incoherent channel, making it improbable for coherent errors to imitate the traversable wormhole signal." On the other hand, they noted that "coherent errors are largely overshadowed by the genuine teleportation signal" \cite{Jafferis}. 
Jafferis et al. contend that despite the presence of coherent errors, the observable effects of the genuine quantum teleportation signal are more prominent. In light of this situation, they assert that they have effectively discerned and differentiated the true teleportation signal from the surrounding background noise.

If Kobrin et al. deemed the learned Hamiltonian problematic, it could pose challenges in differentiating between coherent noise and the genuine quantum teleportation signal. In such a scenario, accurately identifying the true signal from the background noise becomes more complex, and the observable effects become less pronounced and more ambiguous. The ability to distinguish between the two may be compromised due to the issues with the learned Hamiltonian, leading to uncertainties and potential limitations in Jafferis et al.'s findings on quantum teleportation through a holographic wormhole. 

Moreover, during the training process of the Hamiltonian, the system's behavior was approximated using simplified models that cannot fully capture all the complexities of the actual Sycamore processor. Even if the training process relies on noisy simulations rather than idealized models of the Sycamore processor, discrepancies and errors inevitably arise. The problem is that the learned Hamiltonian probably performs well in simulations but fails to capture the true behavior of the Sycamore process due to the noise inherent in the processor not being accounted for during training. Consequently, the trained model might not be as accurate when deployed on the Sycamore because of gate parameter fluctuations and environmental noise. As a result, it could produce inaccurate or inconsistent results on the Sycamore processor.

Thus, due to the presence of noise, we cannot have complete confidence in the reliability of the processor. Quantum gates are susceptible to errors stemming from imperfect control over quantum systems and imprecise calibration. Additionally, quantum measurements can introduce noise and disturb the quantum state, affecting subsequent computations. Achieving high confidence in the Sycamore processor poses significant challenges. Noise introduces uncertainties that limit the reliability of the Sycamore processor \cite{Kalai}.

Considering these challenges and the wormhole experiment using only $9$ qubits without aiming to demonstrate quantum supremacy, opting for a classical simulation could have been preferable. We can observe why using a classical simulation is preferable by applying Ian Hacking's ideas to the wormhole experiment.

According to Hacking, the term "model" can sometimes refer to a material model constructed in a laboratory, essentially a simulation of an experiment. For instance, scientists might find it helpful to build a desktop model to gain clearer insights into a particular phenomenon \cite{Hacking}. 

In the case of the experiment performed by Jafferis et al., they utilized the Sycamore, a superconducting quantum processor consisting of transmons (Josephson junctions). By running the experiment on the $9$ qubits, Jafferis et al. studied how the qubit is transferred from the left side to the right side, thus deepening their understanding of the teleportation-through-a-wormhole phenomenon. Implementing the simulation of teleportation between the left and right sides involves precise control and manipulation of the $9$ qubits (or transmons). It requires carefully designed quantum gates and methods to prepare, entangle, and measure the qubits appropriately.

According to Hacking, using the material model allows scientists to obtain accurate inputs and outputs, leading to a better understanding of the phenomenon and potentially generating ideas for enhancing experiments. However, Hacking emphasizes that such a model is not an exact representation of reality; it is not a literal depiction of how things truly exist \cite{Hacking}.

In other words, the experiment by Jafferis et al. involves running a quantum teleportation simulation using $9$ qubits in a superconducting quantum processor. But it's essential to understand that such a model does not accurately represent reality; namely, Jafferis et al.'s simulation does not literally represent an actual teleportation through a traversable wormhole event. The quantum teleportation process, as simulated in the laboratory, does not mean that physical particles are instantaneously moving between the left and right sides of the processor. A (superconducting) qubit is not a particle. It typically comprises several physical devices and components working together to represent and manipulate quantum information (the Josephson junction\footnote{Which allows the qubit to exhibit quantum coherence and behave as a two-level quantum system.}, microwave pulses to manipulate the qubit's quantum state, etc.). 
In other words, the model serves as a tool for investigation and exploration, but it should not be confused with a genuine manifestation of a traversable wormhole.
 
\section*{Acknowledgement}

\noindent This work is supported by ERC advanced grant number 834735.

\end{document}